\documentclass[11pt]{article}
\usepackage{shuzenji,natbib}
\usepackage{graphicx}
\newcommand{\etal}{et al.}
\newcommand{\eg}{{\it e.g.,}}

\newcommand{\GC}{Galactic Centre}
\newcommand{\GCR}{Galactic Centre Region}
\newcommand{\XMM}{{\it XMM}}
\newcommand{\XMMN}{{\it XMM-Newton}}
\newcommand{\Chandra}{{\it Chandra}}
\newcommand{\LUMIUNIT}{erg~s$^{-1}$}
\newcommand{\NHUNIT}{H~cm$^-2$}
\newcommand{\SgrAEast}{Sgr~A East}

\begin{document}
\vspace*{13mm} \noindent \hspace*{13mm}
\begin{minipage}[t]{14cm}
{\bf THE {\XMMN} VIEW OF THE GALACTIC CENTRE}
\\[13mm]
M. Sakano$^{1,3}$, R. S. Warwick$^{1}$, and A. Decourchelle$^2$\\
{\it $^1$ Dept. of Physics and Astronomy, Univ. of Leicester,
 Leicester LE1 7RH, UK} \\ 
{\it $^2$ CEA/DSM/DAPNIA, Service d'Astrophysique, C.E. Saclay, 
        91191 Gif-sur-Yvette Cedex, France} \\ 
{\it $^3$ Japan Society for the Promotion of Science (JSPS)} \\ 
\end{minipage}

\section*{Abstract}

In X-rays, the Galactic Centre (GC) Region appears as a complex 
of diffuse thermal and non-thermal X-ray emission intermixed with a 
population of luminous discrete sources. Here we
present some new findings from the {\XMMN} GC survey, focusing
particularly on  Sgr~A$^*$, Sgr~A East, the Radio Arc region, and a newly
discovered X-ray non-thermal filament.

\section{Introduction}

Our {\GCR} provides, arguably, the best available laboratory in which 
to study the central activity of galaxies.  Recent measurements of the proper 
motion of infrared stars around the dynamical centre of the Galaxy coupled with
the enhanced radio brightness of the region, strongly support the
presence of a super massive black hole of mass (2--3)$\times 10^6$
M$_\odot$ ({\eg} Genzel {\etal} 2000) in Sgr~A$^*$.  A
number of unusual phenomena are also observed in the {\GCR}: {\eg}
magnetic radio filaments, high density, fast-moving molecular clouds,
massive star clusters, and high-temperature plasma extending over scales
in excess of a few hundred parsecs, etc.

A wide-angle {\XMMN} survey of the {\GCR} was carried out in the
period 2000--2001 utilising 11 pointings in total, each with 10--25 ksec 
exposures (see Warwick 2002).  This {\XMMN} {\GC} survey along with a
complementary {\Chandra} survey (Wang {\etal} 2002) has revealed the 
complex X-ray structure of the region.  Here we report some of the recent
discoveries which have stemmed from the {\XMM} {\GC} Survey, focusing
particularly on Sgr~A$^*$, Sgr~A East, the Radio Arc region, and an
X-ray non-thermal filament designated XMM J174540$-$2904.5.

\section{Global image}

Fig.~\ref{fig:img} shows a false colour X-ray image of the {\GCR} taken
with {\XMMN} in which Galactic north is up.  Most of red dots are
presumably foreground stars.  Two persistent (1E 1740.7$-$2942 and 1E
1743.1$-$2843) and one transient (SAX J1747.0$-$2853) X-ray binaries are
apparent as very bright sources in these observations.  There is  
extended high surface brightness emission coincident with the Sgr~A and 
Radio Arc (or G0.1$-$0.1) regions.  The Arches star cluster and SNR 
G0.9$+$0.1 are also prominent in the hard energy band.  In addition to these
identifiable sources, low-surface brightness emission is evident over the
whole region. The configuration is, however, highly asymmetric with the region 
to the Galactic west of Sgr~A being brighter and having a more complex 
morphology than that to the east.  We also point out that there is an 
extended and clumpy structure extending from Sgr~A towards the (Galactic) 
south-west direction on a  scale of several tens of parsecs.

\section{Sgr~A$^*$}

With the moderate angular resolution of {\XMMN}, Sgr~A$^*$ cannot be
resolved.  However, we have detected an X-ray flare from Sgr~A$^*$.
Since the flare was detected in the last tens of minutes of an
observation and was increasing in brightness right up to the end of the
observation, the peak luminosity is unknown; the maximum luminosity
observed was 4$\times 10^{34}${\LUMIUNIT}.  The flare nature is
basically consistent with those detected by {\Chandra} (Baganoff {\etal} 
2001).  Full details are given in Goldwurm {\etal} (2002).

\section{Sgr~A East}

\subsection{Images}

The Sgr~A region which is characterised by extremely bright radio 
structure, consists of Sgr~A East and West.  Sgr~A East is an oval shell-like 
radio morphology, and in projection surrounds Sgr~A West, which includes 
Sgr~A$^*$.
From its shell-like structure, {\SgrAEast} is supposed to be a supernova
remnant (SNR), SNR 000.0+00.0, although alternative
interpretations have also been proposed.

Fig.~\ref{fig:img-sgraest} shows the {\XMMN} MOS1+2 image of the Sgr~A
region.  The X-ray emitting region is contained
within the radio shell of {\SgrAEast}, as is also readily 
apparent from the {\Chandra} observation (Maeda {\etal} 2002).
We also made a 6.7-keV line narrow-band image corresponding to
helium-like iron, since the X-ray spectrum of {\SgrAEast} apparently 
shows that line (see Sec.\ref{sec:sgraest-spec}).
The underlying continuum is subtracted using an adjacent bandpass and 
assuming an averaged spectral shape (Fig.~\ref{fig:img-sgraest} right panel).
The 6.7-keV line is found to be clearly more concentrated in the core of
{\SgrAEast} than the continuum. 
This implies that the core of {\SgrAEast} is more
abundant in iron or higher in temperature, or perhaps 
a combination of these factors (see Sec.\ref{sec:sgraest-spec} for detail).

\subsection{Spectra \label{sec:sgraest-spec}}

We accumulated the source spectrum of Sgr A East from a region of
radius 100$''$ (see Fig.~\ref{fig:img-sgraest}), but excluding both Sgr~A West
and the region around a bright soft point source.
In the spectrum, several strong
emission lines at energies corresponding to K$\alpha$ lines from 
highly ionized ions, can be seen (Sakano {\etal} 2002b).

From the line ratios of K$\alpha$ lines from helium(He)-like and
hydrogen(H)-like atoms, we found that the spectrum requires a
multi-temperature plasma.  When we approximate the spectrum with a
two-temperature thin thermal plasma model by fitting all the spectra 
(MOS1, 2, and pn)
simultaneously, we obtained the best-fitting temperatures to be
$\sim$1~keV and $\sim$4~keV.
The best-fit abundances for silicon,
sulfur, argon, calcium, and iron vary slightly from element to
element, but are mostly in the range 1--2 solar units.

We further examined the possible spectral variation within
{\SgrAEast} using spatially resolved spectra, and found that only the
iron abundance varies significantly from place to place: $\sim$0.5
solar in the outer region ($r>60''$) to $\sim$3.5 solar in the central
region ($r<28''$).  Thus, the iron line concentration in
Fig.~\ref{fig:img-sgraest} is due to an
abundance gradient.

\subsection{Discussion}

The concentration of the iron at the centre suggests that the plasma
originated in ejecta.
From the derived spectral parameters and spatial extent, we estimated
the total energy and mass to be $1.5\times 10^{49} \eta^{1/2}$ erg and
1.9$\eta^{1/2}$ M$_{\odot}$, respectively, assuming pressure
equilibrium between the low- and high-temperature components, where
$\eta$ is a filling factor of the total plasma.  These results, therefore, 
favour an origin for {\SgrAEast} in a type-II SN event.  The most
intriguing problem is the  extremely high temperature of 4~keV.  The
special environment in the {\GCR} may be responsible for it.
The detailed analysis and discussion are found in Sakano {\etal} (2002b).

\section{The Radio Arc region}

This region is unique due to the bright structure both
in the radio and X-ray bands.  The brightest X-ray emission is coincident
with the position of the molecular cloud G~0.1$-$0.1.  On
the other hand, the radio brightest structure, the Radio Arc, is not
particularly luminous in the X-ray band.  Another unusual
characteristic is the extremely high surface
brightness of the 6.4-keV line, a fluorescent K$\alpha$ line from
neutral iron.  Whereas the X-ray reflection nebula, Sgr~B2, is dim in
the continuum but bright only in the 6.4-keV line (Murakami {\etal}
2001), this region is bright both in the continuum and 6.4-keV line.
The origin of this X-ray structure remains an open question.

\section{An X-ray non-thermal filament XMM J174540$-$2904.5}

\begin{figure}[htb]

  \begin{minipage}{16cm}
     \includegraphics[width=16cm]{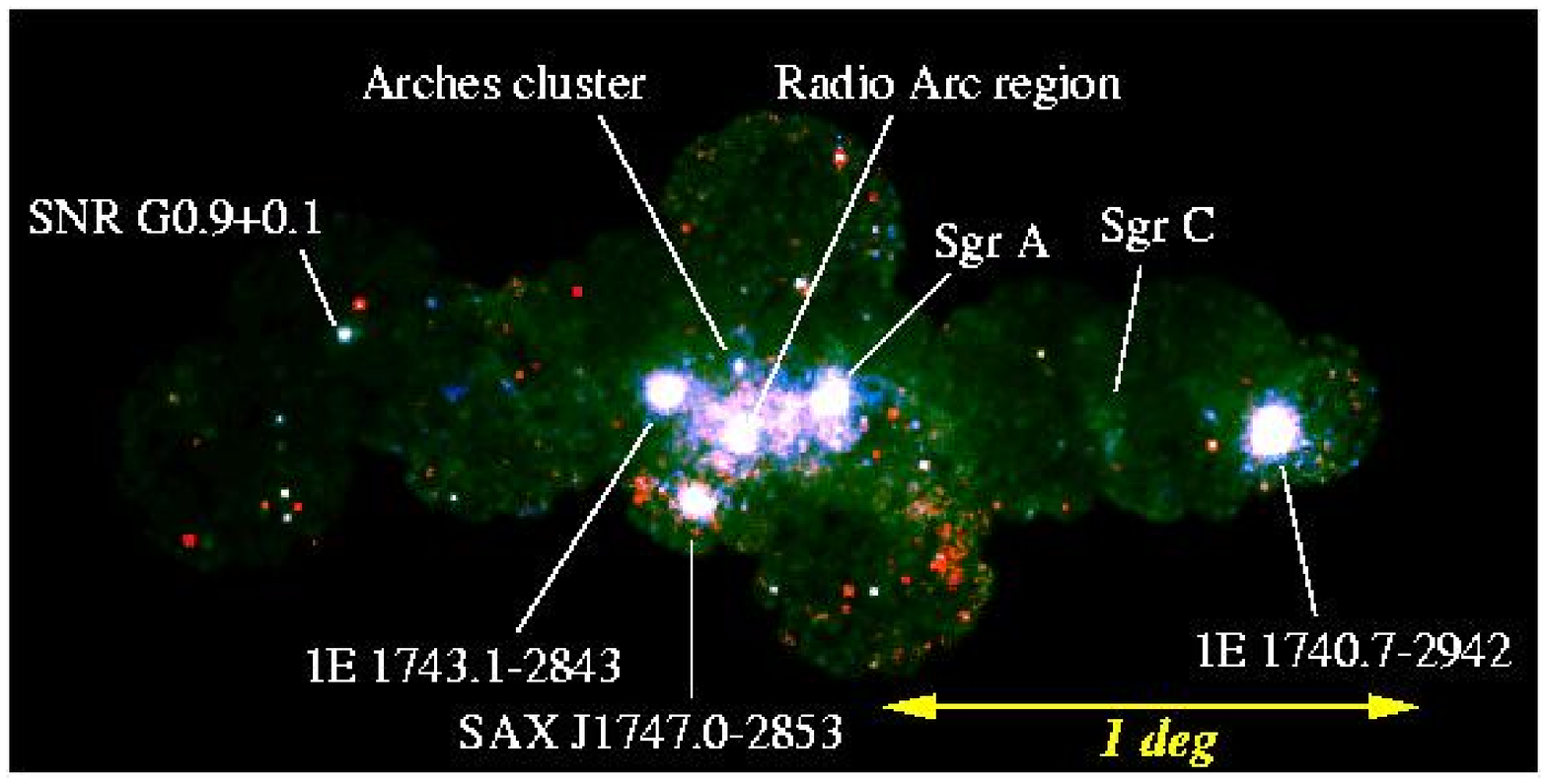}
     \caption{{\XMMN} MOS1+2 colour image of the {\GCR} along the
   Galactic Plane.
   \label{fig:img}
	}
     \includegraphics[width=16cm]{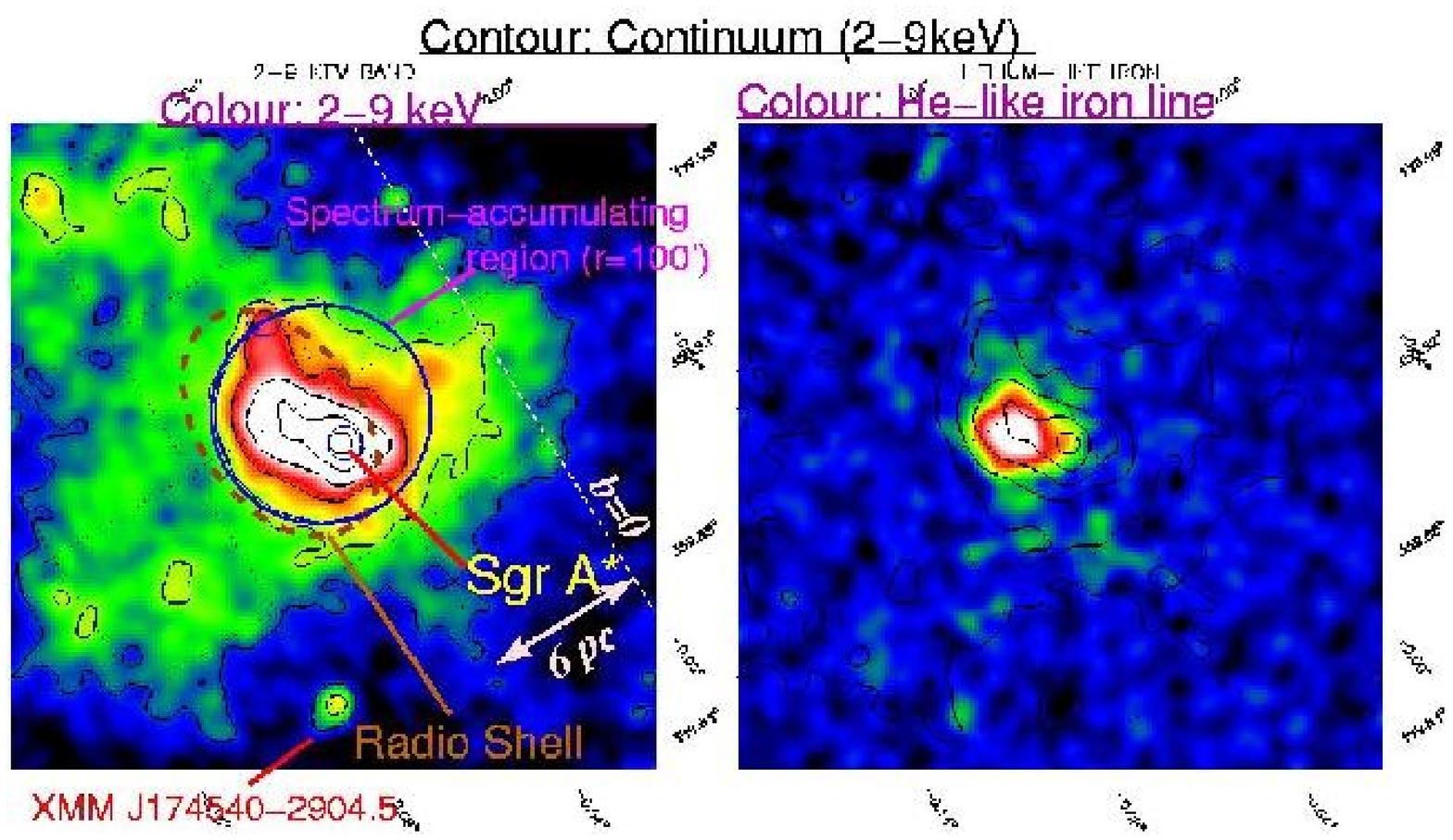}
     \caption{(Left) {\XMMN} MOS1+2 image of the Sgr~A region in the
   2--9~keV band.  (Right) The 6.7-keV He-like iron line image after the
   continuum image is subtracted.  The contour for the 2--9 keV
   continuum is overlaid.  The dashed lines represent the galactic coordinates.
  \label{fig:img-sgraest}
	}
  \end{minipage}
\end{figure}

A hard extended source, XMM J174540$-$2904.5, newly discovered with 
{\XMM}, is marked in Fig.~\ref{fig:img-sgraest}
This source has a corresponding radio non-thermal structure, the Sgr A-E
`wisp' (Ho {\etal} 1985) (= 1LC 359.888$-$0.086 = G359.88$-$0.07).  Both 
the X-ray and radio images show similar elongated structure, although
the X-ray emission is more localised.  We found
that the X-ray emission from this source is clearly non-thermal with an 
energy index of 1.0$^{+1.1}_{-0.9}$ and heavily absorbed ($N_{\rm
H}=38^{+7}_{-11}\times 10^{22}${\NHUNIT}).  So far, this is the first
X-ray/radio filament in the {\GCR} which has been shown, unequivocally,
to have a non-thermal X-ray spectrum.

The spectral energy distribution shows that the X-ray
spectrum is steeper than that observed in the radio band 
but is, most likely, smoothly connected to it.
Thus, the emission is presumably due to synchrotron radiation.  The
lifetime of the high energy particles ($\sim$7~yr) emitting X-rays is
roughly comparable with the spatial extent of the structure.  Hence, the
X-ray source is probably the site of particle acceleration.  The position of
the X-ray source coincides with the peak of the molecular cloud and it
is quite possible that dense environment is one factor in determining this 
as an efficient acceleration site. This represents one of only a few 
X-ray/radio non-thermal filament known in the {\GCR}.  The details are 
discussed in Sakano {\etal} (2002a).

\bigskip M.S. are financially supported by JSPS.

\end{document}